\documentclass[twocolumn,aps,showpacs,amsmath,preprintnumbers,floatfix,a4paper]{revtex4-1}

\usepackage{graphicx}
\usepackage{dcolumn}
\usepackage{bm}
\usepackage{color}
\usepackage{amssymb}
\usepackage{natbib}
\usepackage{hyperref}

\begin{document}
	
\title{Particle resonances and trapping of direct laser acceleration in a laser-plasma channel}

\author{F.-Y. Li}
    \thanks{fyli.acad@gmail.com\\
    Presently at the New Mexico Consortium\\
    Los Alamos, NM 87544}
\author{P. K. Singh}
\author{S. Palaniyappan}
\author{C.-K. Huang}
	\thanks{huangck@lanl.gov}
		
\affiliation
{
Los Alamos National Laboratory, Los Alamos, New Mexico 87545, USA
}
\date{\today}

\begin{abstract}
As one of the leading acceleration mechanisms in laser-driven underdense plasmas, direct laser acceleration (DLA) is capable of producing high-energy-density electron beams in a plasma channel for many applications. However, the mechanism relies on highly nonlinear particle-laser resonances, rendering its modeling and control to be very challenging. Here, we report on novel physics of the particle resonances and, based on that, define a potential path toward more controlled DLA. Key findings are acquired by treating the electron propagation angle independently within a comprehensive model. This approach uncovers the complete particle resonances over broad propagation angles, the physical regimes under which paraxial/non-paraxial dynamics dominates, a unified picture for different harmonics, and crucially, the physical accessibility to these particle resonances. These new insights can have important implications where we address the basic issue of particle trapping as an example. We show how the uncovered trapping parameter space can lead to better acceleration control. More implications for the development of this basic type of acceleration are discussed.    
\end{abstract}


\maketitle

\section{Introduction}
Utilizing high-power lasers for high-energy electron acceleration in plasmas has been intensely pursued in the last few decades~\cite{mourou2006optics,esarey2009physics}. Two major types of acceleration have been exploited: direct acceleration by the laser fields (i.e., direct laser acceleration or DLA)~\cite{pukhov1999particle} and indirect plasma-field acceleration induced in the laser wake (i.e., laser wakefield acceleration)~\cite{tajima1979laser}. 
Both scenarios operate most efficiently when the laser drives a plasma channel by expelling electrons outward. The wakefield regime, nevertheless, favors an ultrashort femtosecond laser driver, where trapped electrons are separated from the driving pulse and experience a longitudinal plasma acceleration along the laser propagation direction~\cite{tajima1979laser}. Tremendous efforts have been devoted to controlling electron trapping in the wakefield with suitable laser-plasma conditions~\cite{esarey1997electron,pak2010injection,gonsalves2011tunable}, which have led to high-quality generation of pC-charge GeV electron beams~\cite{gonsalves2019petawatt}.

In this paper, the other major type of acceleration, i.e., DLA, is considered with an end goal of improving its control. DLA typically occurs with a long picosecond driving laser, which creates an extended channel with fields dominated by the transverse component, enforcing transverse betatron oscillations. Meanwhile, trapped electrons are subject to the overlapping laser fields, such that DLA is invoked by particle-laser resonances when the betatron oscillation matches witnessed laser oscillation~\cite{pukhov1999particle}. This process represents a strong laser-electron coupling and produces high-current electron beams of enormous nC-$\mu$C charge~\cite{rosmej2020high}, which can drive ion acceleration and high-dose x/$\gamma$-rays, neutrons, and positrons for medical, nuclear, and radiography applications~\cite{kneip2008observation,liu2015quasimonoenergetic,chen2015scaling,fernandez2017laser,bin2018enhanced}. Moreover, DLA works for a wide range of plasma densities due to relativistic and preplasma effects~\cite{gahn1999multi,liu2013generating,palaniyappan2012dynamics,tsymbalov2019well}. Despite its importance, experiments concerning the DLA process often observe poor beam quality and, sometimes, a low generation efficiency~\cite{gahn1999multi,mangles2005electron,rosmej2020high}.

Deeper insights into the DLA physics are much needed in order to improve its performance. Many studies using Hamiltonian analysis~\cite{zhang2018electron}, and Monte-Carlo~\cite{tsakiris2000laser} and particle-in-cell simulations~\cite{pukhov1999particle,liu2013generating} have been reported. Existing understandings are, however, limited and mixed even regarding a few fundamentals. 
The paraxial approximation of electron propagation has been adopted in all previous analysis, assuming but with no justification that the electron transverse-to-longitudinal momentum ratio satisfies $\xi_1\equiv p_y/p_x\ll 1$. Moreover, only first-order resonance (i.e., betatron and laser frequencies match exactly) has been typically considered~\cite{pukhov1999particle,khudik2016universal}, while very high-order ones (i.e., laser frequency being multiple that of betatron) are interpreted qualitatively differently as a stochastic effect~\cite{zhang2018stochastic}. 
As a consequence, basic questions like whether non-paraxial dynamics exists, what physical regimes each refers to, how the first and high-order resonances are correlated, and crucially, what determines their accessibility remain unclear. 
These gaps have left some basic elements for controlling DLA yet to be well defined. Of particular relevance here (and also to any advanced accelerator concepts) is a better appreciation of particle trapping, such that controlled acceleration is possible by tailoring the laser-plasma conditions.

The central new result reported in this paper is a novel framework to address the above basic questions and to define a solid step toward controlled DLA.   
A model incorporating major features of DLA is first proposed, which allows for a formal introduction of the resonances at arbitrary order. The key novelty then is to treat the electron propagation angle explicitly and let it vary independently from other dynamics. This approach is found to give rich new physics that the usual paraxial assumption misses and are essential for establishing some fundamental aspects of DLA. We apply the results to particle trapping and show how better design of DLA may be pursued based on the uncovered trapping parameter space.

\section{Physical model}
We start with a two-dimensional (2D) model of DLA.  The channel focusing field is given by $E_{y,C}=k_ey$ and the laser fields by $E_{y,L}=a_0\cos\phi$, $B_{z,L}=E_{y,L}/v_p$, where $k_e=\omega_p^2/2\omega_0^2$, $\omega_p$ is the plasma frequency, $\omega_0$ the laser frequency, $a_0$ the normalized laser amplitude, $\phi=t-x/v_p$ the laser phase, and $v_p$ the phase velocity. Here, the fields are normalized to $m_e\omega_0c/e$, space to $c/\omega_0$, time to $1/\omega_0$, and velocities to the light speed $c$. The model can be extended to a Gaussian laser~\cite{tsakiris2000laser} and 3D channel of electromagnetic fields~\cite{pukhov1999particle,huang2017nonlinear} with our key results unaffected. The electron dynamics follows the relativistic equations of motion (REM) $d\vec{\textbf{p}}/dt=-\vec{\textbf{E}}-\vec{\textbf{v}}\times\vec{\textbf{B}}$, $d\gamma/dt=-\vec{\textbf{v}}\cdot\vec{\textbf{E}}$, $d\vec{\textbf{r}}/dt=\vec{\textbf{p}}/\gamma$, with momentum $\vec{\textbf{p}}$ normalized to $m_ec$ and $\gamma$ the relativistic Lorentz factor.

Despite high nonlinearity the system has a constant of motion (CoM) $\gamma-p_x+k_ey^2/2=C_0+f(p_x)$~\cite{pukhov1999particle}, where $C_0=(1+p_{x0}^2+p_{y0}^2)^{1/2}-v_pp_{x0}+k_ey_0^2/2$ is related to initial injection parameters $(p_{x0}, p_{y0}, y_0)$ and $f(p_x)=\epsilon p_x=(v_p-1)p_x$.
The CoM importantly implies bounded betatron oscillations~\cite{arefiev2012parametric}, i.e., $y_{\rm max}\to y_b\equiv (2C_0/k_e)^{1/2}$, because the term $\gamma-p_x=(1+p_x^2)^{1/2}-p_x\equiv g(p_x)$ nearly reduces to zero at the oscillation boundaries where $p_y=0$ and $p_x\gg 1$. 
With the regularized betatron motion $y=hy_b\cos\theta$ (where $h\equiv [1+\frac{f(p_x)}{C_0}-\frac{g(p_x)}{C_0}]^{1/2}\to 1)$, we simplify the REM by transforming to the frame of the betatron phase $\theta$,
\begin{subequations}
    \label{Gamma_phi_ODEs}
    \begin{align}
        \label{Gamma_phi_ODEs1}
        d\gamma/d\theta&=a_0hy_b\sin\theta\cos\phi+C_0h^2\sin2\theta,\\
        \label{Gamma_phi_ODEs2}
        d\phi/d\theta&=C_0(1-h^2\cos^2\theta)/\sqrt{k_e\gamma}.
    \end{align}
\end{subequations}
It is seen that DLA or the laser work, $W_{L}=\int a_0hy_b\sin\theta\cos\phi d\theta$, depends on the beating of the laser phase ($\phi$) with the betatron phase ($\theta$). The integral is simplified by considering laser perturbation to a large-amplitude betatron oscillation, which leads to the phase-matching condition, $\phi=l\theta+d_1\sin2\theta+\phi''$, where $l$, $d_1$ and $\phi''$ are constants related to initial parameters (see the Appendix for derivation). Plugging it back to $W_L$, we find that odd numbers of $l$ or odd-harmonic resonances are required for DLA (or $W_L$) to be pronounced over multiple betatron cycles. Therefore, the general frequency-matching condition (FMC) is obtained as $d\phi/dt=ld\theta/dt$ by dropping the small term $\sin2\theta$. Making use of the betatron frequency $d\theta/dt=\sqrt{k_e/\gamma}$ and witnessed laser frequency $d\phi/dt=1-p_x/\gamma v_p$~\cite{pukhov1999particle}, we cast the FMC as
\begin{equation}
\label{frequency_matching}
l\sqrt{k_e\langle\gamma\rangle}=\langle\gamma-p_x\rangle+\chi \langle f(p_x)\rangle; l=1,3,5, ...,
\end{equation}
where $\chi=1/v_p$ and $\langle ... \rangle$ refers to averaging over the betatron phase; we hereafter omit the averaging symbol for simplicity.

\begin{figure}[t]
	\centering
	\includegraphics[width=0.48\textwidth]{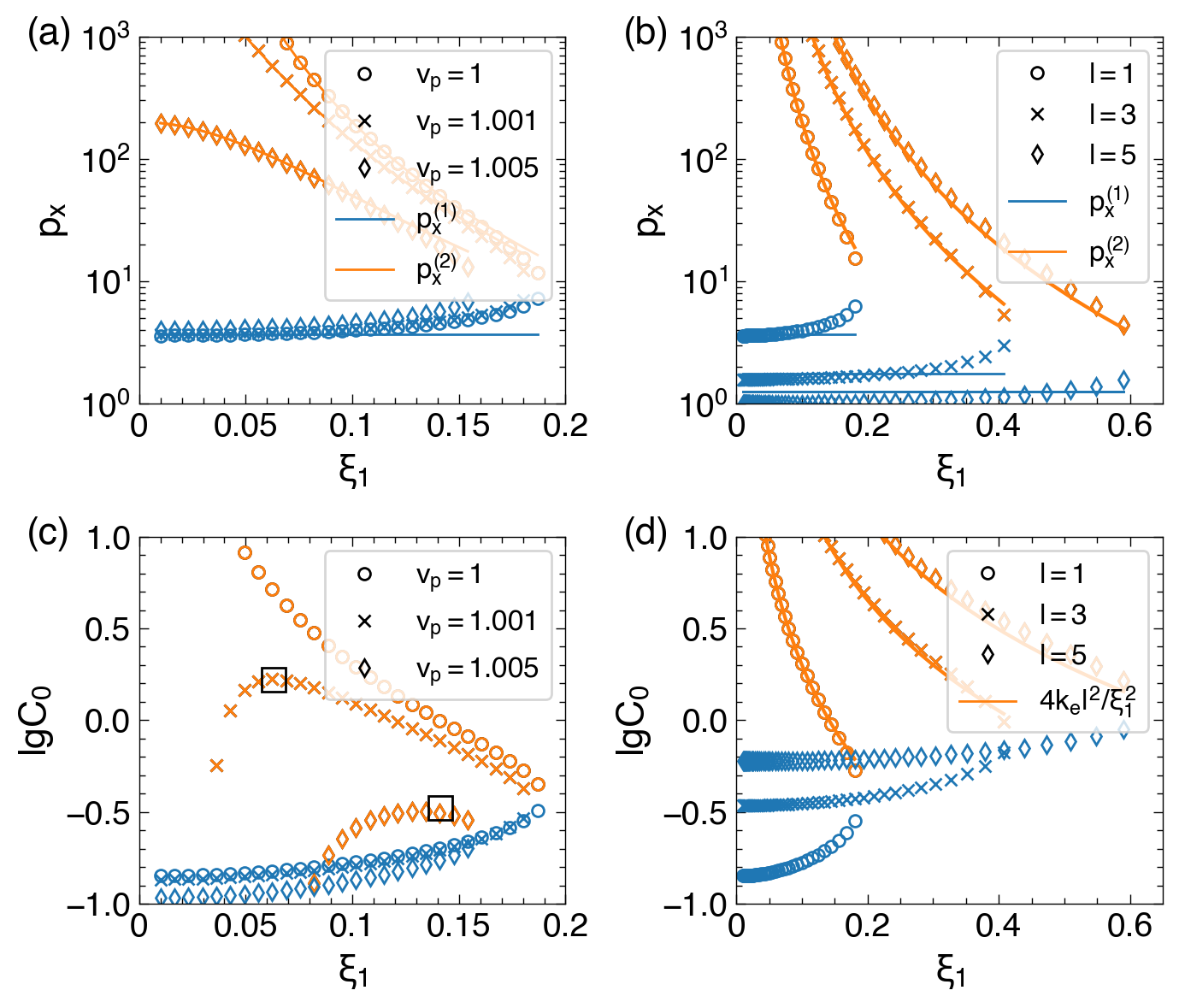}
	\caption{(a) Resonance $p_x$ versus $\xi_1$ for $l=1$ and varying $v_p$ and (b) for $v_p=1$ and different $l$. (c,d) Constant $C_0$ versus $\xi_1$ at corresponding conditions of (a,b). The blue and orange colors represent the low and high branches, respectively. The solid lines refer to Eq.~(\ref{res_limits}) in (a,b) and Eq.~(\ref{mapping}) in (d). In the calculations we take $\omega_p/\omega_0=0.1$.
	}
	\label{resonance_solution}
\end{figure}

\section{Complete solution of particle resonances and their accessibility}
Let us first solve the momentum $p_x$ as required for achieving DLA resonances. To retain the propagation angle we cast $\gamma= p_x (1+\xi_1^2+\xi_2^2)^{1/2}$ with $\xi_2=1/p_x$, and crucially let $\xi_1=p_y/p_x$ be independent from $p_x$. This way the FMC is arranged in $\xi_2$, or $p_x$, as $(1-k_e^2l^4)\xi_2^4-4k_e\chi l^2\xi_2^3+[(2-l^4k_e^2)(1+\xi_1^2)-2\chi^2]\xi_2^2-4k_e\chi l^2(1+\xi_1^2)\xi_2+[\xi_1^4+(1-\chi^2)(1-\chi^2+2\xi_1^2)]=0$. This polynomial is found to give a pair of real roots for a wide range of physically allowed $k_e$, $v_p$ and $\xi_1$. Figures~\ref{resonance_solution}(a,b) display the solved $p_x$ against all allowed $\xi_1$ under different laser phase velocities and harmonic orders. Here we uncover for the first time that there exists a natural spread of $\xi_1$ that hosts distinct two-branch DLA resonances (marked as different colors) before they become degenerated at some cutoff. This result is contrary to usual perceptions of $p_y$ being much smaller than $p_x$; instead it tells that $p_y$ could be proportionally larger as $p_x$ increases in DLA.

To clarify the solution structures and understand involved physics, we extract $p_x=4k_ev_p^2l^2/(\xi_1^2+\xi_2^2+2\epsilon)^2$ from the FMC under $\xi_1^2\ll 1$.
It is further simplified under the limits of (1) $v_p=1, \xi_1=0$ and (2)  $\xi_1\gg\xi_2$ as
\begin{equation}
\label{res_limits}
p_x^{(1)}=1/(4k_el^2)^{1/3}, \hspace{0.1in} p_x^{(2)}=4k_el^2/(\xi_1^2+2\epsilon)^2.
\end{equation}
Appended as solid curves in Figs.~\ref{resonance_solution}(a,b), it is seen that the two limits correspond to the low and high branches, respectively.
The actual low branch at finite $\xi_1$ [blue points in Figs. 1(a-b)] only increases slightly from the limit $p_x^{(1)}$ obtained at $\xi_1=0$. The high branch (orange points), despite being infinitely large at $\xi_1=\epsilon=0$, drops quickly with $\epsilon$ or $\xi_1$. In particular, $p_x^{(2)}$ scales with $\xi_1$ as $1/\xi_1^4$, thus it becomes low enough to be more accessible at larger $\xi_1$.  
Motivated by the distinct trends, the cutoff in $\xi_1$ is estimated as
\begin{equation}
\label{cutoff}
    \xi_1^{\rm cut}\simeq (4k_el^2)^{1/3}=2^{1/3}(\omega_p/\omega_0)^{2/3}l^{2/3},
\end{equation}
by letting $p_x^{(2)}=p_x^{(1)}$. 
This formula very importantly reveals the physical parameter regimes that one should expect for particle resonances in DLA. It clarifies that the paraxial assumption only applies to relatively low-density ($\propto \omega_p^2/\omega_0^2$) and low-harmonic regimes. While for high-density or high-harmonic regime, non-paraxial dynamics may emerge, i.e., $\xi_1\sim\mathcal{O}(1)$. It is worth noting that as the interaction becomes more non-paraxial, the low branch is largely suppressed since we have $p_x^{(1)}\simeq 1/\xi_1^{\rm cut}$. The suppression can be seen in Fig.~\ref{resonance_solution}(b) as $l$ increases.

Given the broad $\xi_1$-distribution, a natural question next is on the accessibility to each $\xi_1$ for given electron injection. To gain insight into that, we consider the CoM which relates full dynamics to initial injection conditions. Making use of the full betatron amplitude, $y_{\rm max}=hy_b$, we arrive at the following averaged form of the CoM,
\begin{equation}
\label{mapping}
C_0(\xi_1)=2p_x(\sqrt{1+\xi_1^2+\xi_2^2}-1)-f(p_x)-g(p_x).
\end{equation}
It shows a single dependence of $C_0$ on $\xi_1$ by substituting $p_x$ on the right hand side with the above solution of $p_x(\xi_1)$. This relation thus concludes that the accessibility to a particular $\xi_1$ is precisely determined by the injection parameters $(p_{x0}, p_{y0}, y_0)$ grouped as the single constant $C_0$. 
As presented in Figs.~\ref{resonance_solution}(c,d),
two distinct branches are again found for $C_0(\xi_1)$ which is a direct outcome of the peculiar  $p_x(\xi_1)$ structure. Making use of $p_x^{(2)}$ at $v_p=1$, the high-branch $C_0$ has a simple scaling of $C_0=\frac{4k_el^2}{\xi_1^2}$ as shown by the line plots in Fig.~\ref{resonance_solution}(d).

\begin{figure}[t]
	\centering
	\includegraphics[width=0.48\textwidth]{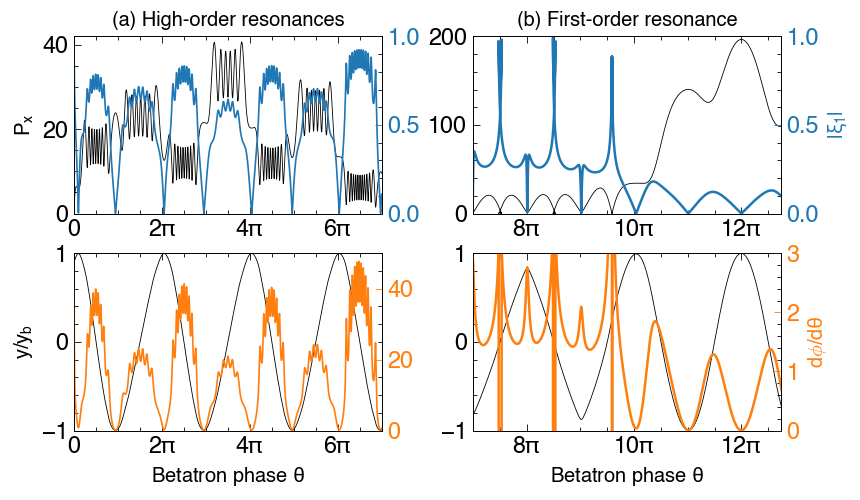}
	\caption{Electron dynamics ($p_x, y, |\xi_1|, d\phi/d\theta$) against betatron phase $\theta$ for panel (a) high-order resonances with $a_0=2, p_{x0}=2, p_{y0}=12, C_0=10.2$ and for (b) first-order resonance with $a_0=4, p_{x0}=2, p_{y0}=2, C_0=1$, where $v_p=1, \omega_p/\omega_0=0.1$.}
	\label{elec_dyn}
\end{figure}

The above mapping of ($p_x, C_0$) versus $\xi_1$ until a well-defined $\xi_1^{\rm cut}$ forms our central new insight into the DLA dynamics. It describes clear physics that an injection characterized by $C_0$ potentially leads to DLA only at a particular set of resonance $\xi_1$ and $p_x$ achieved under different harmonics. By eliminating the $\xi_1$ dependence we arrive at a direct mapping between $p_x$ and $C_0$, e.g., $p_x=\frac{C_0^2}{4k_el^2}$ for the high branch and $v_p=1$.
It implies $p_x$-transition between different harmonics, a phenomenon interpreted as a stochastic effect only for $l\gg 1$~\cite{zhang2018stochastic}. It also implies $\xi_1$-transition following $\xi_1=(\frac{4k_el^2}{C_0})^{1/2}$. These dynamics can be more clearly seen in the top-panel of Fig.~\ref{elec_dyn}(a), where the REM are directly integrated for certain initial injection conditions corresponding to high harmonic resonances. 

Notice that, for $v_p>1$ which may happen for incomplete channel evacuation, the high branch $C_0$ is capped by $C_0\leq\frac{k_el^2}{3\epsilon}$ with the peak location $\xi_1\simeq 2\epsilon^{1/2}$ independent of the harmonic order $l$ [squares in Fig.~\ref{resonance_solution}(c)]. This feature only constrains the allowed harmonic orders (i.e., $l\geq\sqrt{\frac{3\epsilon C_0}{k_e}}$) for given injection $C_0$ and does not qualitatively change the above physical picture identified under $v_p=1$.

The mapping relations naturally unify both paraxial/non-paraxial and first/high-harmonic regimes by using the single parameter $\xi_1$. This capability is unique as it opens the possibility to treat extensive electron parameter space under a single framework. It is particularly important because, as we shall see, each individual electron may participate in the nonlinear DLA dynamics differently and even the same particle may behave differently under different laser-plasma conditions. It also points to the strong limitations of the usual paraxial approach which inherently misses the above physics.

\section{Particle trapping condition}
\begin{figure}[t]
	\centering
	\includegraphics[width=0.48\textwidth]{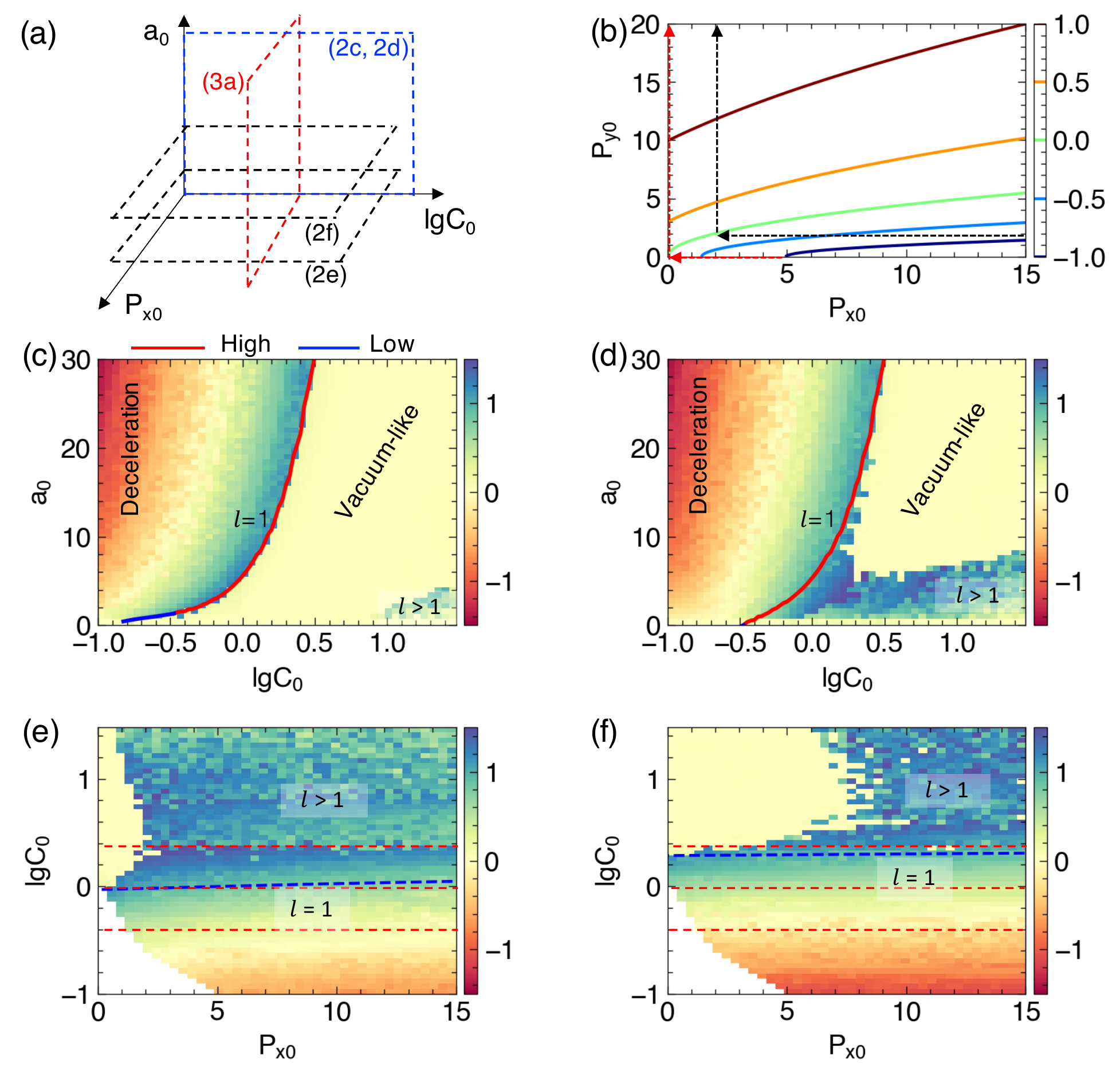}
	\caption{(a) Sketch of 3D trapping space ($a_0, \lg C_0, p_{x0}$). (b) $C_0$ spanned by $p_{x0}, p_{y0}$ ($y_0=0$) with lines displaying constant $\lg C_0$. (c-f) $\lg(\gamma_{\rm m}/\gamma_{\rm m}^{\rm vac})$-distribution in 2D cuts sketched in (a), where each pixel corresponds to a test-particle simulation with implied initial conditions and $v_p=1, \omega_p/\omega_0=0.1$. The red/blue lines in (c,d) represent Eq.~(\ref{a0_required}) for the high/low-branch, respectively. (e,f) correspond to $a_0=5, 15$, respectively, where the blue dashed lines correspond to Eq.~(\ref{px0_bound}). 
	}
	\label{mapping_verification}
\end{figure}

Here we apply these new insights to tackle the problem of particle trapping, which is of fundamental importance to DLA control. Unlike in most linear acceleration schemes (e.g., the wakefield scenario), particle trapping has been poorly characterized in DLA due to its high nonlinearity. Trapping or the onset of DLA essentially relies on small electron-laser dephasing~\cite{arefiev2012parametric}, such that the electron sees more synchronized laser fields. Our model [Eqs.~(\ref{Gamma_phi_ODEs},\ref{frequency_matching})] concludes that while nearly zero dephasing is reached at oscillation boundaries, the average dephasing rate is simply equal to the harmonic order $l$, i.e., 
\begin{equation}
	\label{dephasingave}
		\left \langle\frac{d\phi}{d\theta}\right\rangle=\left\langle\frac{C_0(1-h^2\cos^2\theta)}{\sqrt{k_e\gamma}}\right\rangle=l.
\end{equation}
Therefore, high-order resonances having large average dephasing are triggered by electrons naturally hitting transverse boundaries where $\theta=N\pi$ and the dephasing rate reduces to zero; this can be seen by comparing the two subplots of Fig.~\ref{elec_dyn}(a) for example.  However, this same does not apply to the first-order resonance e.g., Fig.~\ref{elec_dyn}(b), which has overall small dephasing and almost continuous energy exchange. As such, it has to be invoked by a strong acceleration near its onset, i.e., $\Delta p_x\sim v_ya_0\Delta t\geq p_x-p_{x0}$. In terms of resonance quantities, i.e., $v_y\sim \xi_1$, $\Delta t\sim \omega_\beta^{-1}\sim\sqrt{p_x/k_e}$, this requirement can be cast into a threshold for $a_0$,
\begin{equation}
    \label{a0_required}
a_0(C_0, p_{x0})\geq\frac{\mu}{\sqrt{2}}\frac{p_x(C_0)-p_{x0}}{\sqrt{p_x(C_0)}\xi_1(C_0)}\frac{\omega_p}{\omega_0},
\end{equation}  
where $\mu$ is a scaling factor and we have provided the mappings of $p_x(C_0)$ and $\xi_1(C_0)$ to get the dependence on $C_0$. The appearance of $p_{x0}$ shows that the trapping parameter space is essentially 3D ($a_0, C_0, p_{x0})$ [Fig.~\ref{mapping_verification}(a)] instead of 2D ($a_0, C_0$)~\cite{khudik2016universal}. This is caused by the non-uniqueness of ($p_{x0}, p_{y0}$) for given $C_0$ as illustrated by Fig.~\ref{mapping_verification}(b). As we shall see, appreciating this extra dimension is only possible with our inclusive framework.

To first see how well our mappings can be used to describe the trapping threshold, Figures~\ref{mapping_verification}(c,d) present the maximum-energy-ratio [$\lg(\gamma_{\rm m}/\gamma_{\rm m}^{\rm vac})$] distribution in the 2D cut ($a_0, C_0$) of the full 3D space, where $\gamma_{\rm m}^{\rm vac}$ refers to the maximum from vacuum acceleration~\cite{meyer2001relativistic}. The results are obtained from test-particle simulations by directly integrating the original REM over picosecond timescales. Despite the two subplots having the same $C_0$ axis, they correspond to different $(p_{x0}, p_{y0})$ variations following the red and black arrows in Fig.~\ref{mapping_verification}(b), respectively. 
For the first time, we show that the trapping space is divided into first and high-order resonances, as well as deceleration ($\gamma_{\rm m}/\gamma_{\rm m}^{\rm vac}<1$) and vacuum-like dynamics ($\gamma_{\rm m}/\gamma_{\rm m}^{\rm vac}\simeq 1$); see the labels therein. Remarkably, the boundary for the onset of the first-order resonance is well described by Eq.~(\ref{a0_required}) [solid lines], proving the effectiveness of our mapping relations. In particular with case Fig.~\ref{mapping_verification}(c) it consists of segments contributed by the high and low-branch resonances (different line colors), respectively. The low-branch contribution with Fig.~\ref{mapping_verification}(d) is suppressed because the involved $p_{x0}$ [black arrow in Fig.~\ref{mapping_verification}(b)] can be even greater than the low-branch $p_x$.

On the other hand, high-order resonances are triggered at very small laser amplitudes, but suppressed when the laser is strong. The former can be understood from the perturbation regime (i.e., strong betatron oscillation and weak laser) where DLA happens as electrons naturally hit boundaries (where the dephasing vanishes). The suppression at large $a_0$, however, corresponds to strongly non-perturbative dynamics and needs a separate study. Nevertheless, a qualitative interpretation is that the predicted momentum transition gap, $\Delta  p_x|_{l+2}^l=\frac{C_0^2}{k_e}\frac{l+1}{l^2(l+2)^2}$, for the last few low orders becomes so large that the required strong laser effectively turns the dynamics into vacuum-like before transitioning into the first-order resonance.

\begin{figure}[t]
	\centering
	\includegraphics[width=0.48\textwidth]{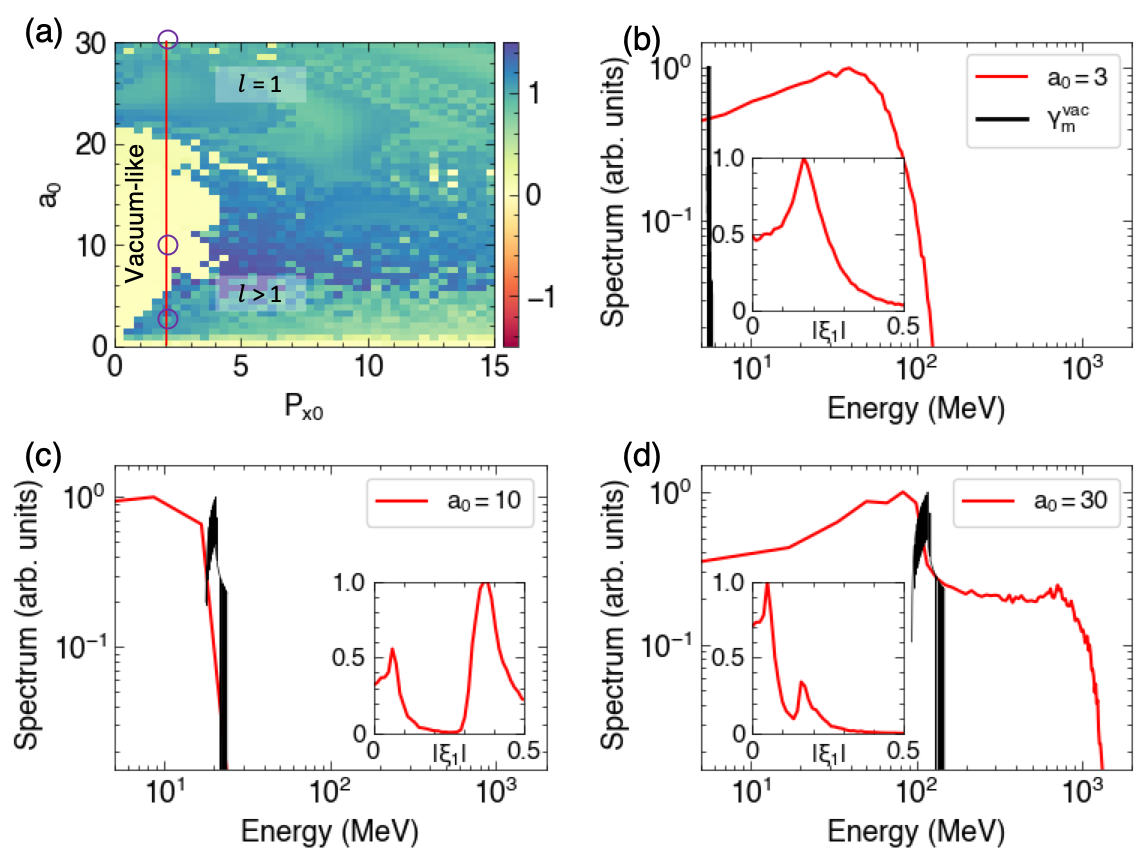}
	\caption{(a) $\lg(\gamma_{\rm m}/\gamma_{\rm m}^{\rm vac})$ distribution in the 2D cut at $\lg C_0=0.4$ as sketched in Fig.~\ref{mapping_verification}(a). (b-d) Energy spectra of a group of side injection [with initial conditions depicted by the red line in (a)] at varying laser amplitudes [marked in (a)]. The insets show corresponding $|\xi_1|$ distributions. The black curves show the distribution of $\gamma_{\rm m}^{\rm vac}$.  
	}
	\label{control}
\end{figure}

To fully appreciate the 3D trapping space, we show in Figs.~\ref{mapping_verification}(e,f) the same $\lg(\gamma_{\rm m}/\gamma_{\rm m}^{\rm vac})$-distribution but along 2D cuts ($\lg C_0, p_{x0}$) at $a_0=5, 15$, respectively. In this space of constant $a_0$, the boundary of first-order resonance can be obtained from Eq.~(\ref{a0_required}) as
\begin{equation}
    \label{px0_bound}
p_{x0}\geq C_0^2/4k_e-(\sqrt{2}/\mu)a_0C_0^{1/2}(\omega_0/\omega_p).
\end{equation}  
Shown as blue dashed lines in Figs.~\ref{mapping_verification}(e,f), they agree reasonably well with the test-particle results, proving the effectiveness of Eq.~(\ref{a0_required}) in describing the general 3D trapping space. Again there are high-order and vacuum-like dynamics found beyond the boundary and affected by the laser amplitude. Since both $C_0$ and $p_{x0}$ are involved, this 2D space is particularly useful for describing general electron parameters. We consider three representative groups of electrons with $\lg C_0$ centered at -0.4, 0, 0.4 (marked as red dashed lines). They may correspond to realistic injection scenarios: (1) $\lg C_0=-0.4$ for pre-acceleration or external injection with large $p_{x0}$~\cite{zhang2015synergistic}; (2) $\lg C_0=0$ for electrons initially at rest, and (3) $\lg C_0=0.4$ for side injection with large ($p_{y0}, y_0$)~\cite{pukhov1999particle}. General dynamics may be inferred by comparing Figs.~\ref{mapping_verification}(e,f). For group (1), DLA functions only at small $a_0$ by the first-order resonance, and large $(a_0, p_{x0})$ only results in deceleration. For group (2), a low $a_0$ threshold exists for DLA to happen. For group (3), DLA is due to high-order resonances at small $a_0$ but more first-order relevant as $a_0$ increases, where a wide vacuum-like dynamics sets the two regimes apart.

To clearly see how the uncovered trapping space may help DLA design, we demonstrate a concrete example with the side injection. It may happen as electrons are first expelled radially during channel formation and then attracted back by plasma oscillation~\cite{pukhov1999particle}. We simulate 50000 electrons in a channel of $\omega_p/\omega_0=0.1$ or ambient density $1.1\times 10^{19} \rm cm^{-3}$ for typical $1\mu m$ laser wavelength. Their initial transverse energy and momenta satisfy $1.7\leq\frac{\mathcal{E}_\perp}{m_ec^2} = \frac{p_{y0}^2}{2\gamma} + \frac{1}{2}k_e y_0^2\leq 2.2, p_{x0}=2, |p_{y0}|>3.2$. The resulting $\lg C_0$ peaks at 0.4 with a spread of 0.1. The 2D trapping space in ($a_0, p_{x0}$), shown at corresponding $\lg C_0$ in Fig.~\ref{control}(a), suggests high-order resonances at $a_0<5$, first-order resonance at $a_0>20$, and vacuum-like dynamics in between. To check their impact on the acceleration, we show the energy and angular ($|\xi_1|$) distributions of the side-injection in Figs.~\ref{control}(b-d) for $a_0=3, 10, 30$, respectively. The distributions are shown at t=6.6 ps, sufficiently long to make their profile stabilize. 
Corresponding $\gamma_{\rm m}^{\rm vac}$-distributions (black curves) are also shown for comparison. It is seen that high-order resonances having lowest threshold $a_0$ generally gives highest acceleration efficiency [Fig.~\ref{control}(b)], with the bulk being accelerated to many times of $\gamma_{\rm m}^{\rm vac}$. This may favor high-yield x-ray generation. The first-order resonance generally results in highest energy cutoff and smallest divergence [Fig.~\ref{control}(d)], but also requires very high-intensity lasers. One should avoid the vacuum-like dynamics where few electrons get accelerated beyond $\gamma_{\rm m}^{\rm vac}$ [Fig.~\ref{control}(c)]. These initial insights show the potential of controlling DLA by matching laser-plasma parameters with possible injection conditions. Much progress on plasma diagnostics has been made recently to enable such tunability~\cite{downer2018diagnostics}

\section{Conclusion}
In conclusion, we have identified important DLA physics that the long adopted paraxial assumption failed to capture, including the full particle resonances and their mapping with the injection conditions. These new insights must be deployed in order to characterize the full electron parameter space because of the high nonlinearity involved in DLA dynamics. We have focused on application to the fundamental issue of particle trapping and showed how better DLA design can be made with the knowledge of the 3D trapping space. Thus our work opens the possibility of controlled trapping in DLA by matching the trapping space with suitable laser-plasma conditions. More developments could be related to analytical beam modeling (e.g., beam divergence) and energy gains, using the present single framework of single parameter $\xi_1$. 
Finally, the insights acquired here may also hint at other devices involving similar acceleration field configurations, such as the free-electron-lasers (FELs)~\cite{huang2007review}, inverse FEL accelerators~\cite{dunning2013demonstration}, and structured interactions~\cite{wang2020power}, especially in their highly nonlinear regime.

\section{Acknowlegments}
F.Y.L. and C.K.H. acknowledge Joshua Burby, Nathan Garland, and Xianzhu Tang for helpful discussions. The work is supported by the Laboratory Directed Research and Development Program of Los Alamos National Laboratory (LANL) under the project 20190124ER. This research used resources provided by the LANL Institutional Computing Program, which is supported by the U.S. Department of Energy National Nuclear Security Administration under Contract No. 89233218CNA000001. 

\section{Appendix: Effective orders of harmonic resonances}
The effectiveness of DLA can be evaluated by the integral, $W_{L}=\int a_0hy_b\sin\theta\cos\phi d\theta$, to see under what conditions the laser work is substantial. The key idea is to perturb from a strong betatron motion by a small-amplitude laser.  For the sake of convenience, we slightly rearrange Eqs.~(\ref{Gamma_phi_ODEs}) as
\begin{subequations}
	\label{odes}
	\begin{align}
		\label{odes1}
		\frac{d \Gamma}{d \theta}&=\frac{1}{2\Gamma} (a_0h y_b \sin\theta \cos\phi + C_0h^2\sin2\theta),\\
		\label{odes2}
		\frac{d\phi}{d \theta}&=\frac{C_0(1-h^2\cos^2\theta)}{\Gamma\sqrt{k_e}},
	\end{align}
\end{subequations}
by letting $\Gamma=\sqrt{\gamma}$. The perturbation method involves three steps: (1) obtain the unperturbed energy variation $\Gamma_0$ due to the betatron motion only, ignoring the laser work or the first term on the RHS of Eq.~(\ref{odes1}); (2) substitute $\Gamma_0$ into Eq.~(\ref{odes2}) to get corresponding witnessed laser phase $\phi$; (3) plug the laser phase into the first term on the RHS of Eq.~(\ref{odes1}) to evaluate the laser work $W_{L}$. 

\subsection{Step 1: Unperturbed energy variation due to betatron motion}
The unperturbed energy variation solely can be obtained by directly integrating Eq.~(\ref{odes1}) and ignoring the first term on the RHS as
\begin{equation}
	\label{Gamma0}
	\Gamma^0 = \sqrt{(\Gamma_0^0)^2- C_0 h^2 \cos^2\theta},
\end{equation}
where $\Gamma_0^0$ is the value of $\Gamma_0$ upon injection, i.e., $\Gamma_0^0=\Gamma^0(\theta=\theta_0)$, $\theta_0=-\pi/2$. 

\subsection{Step 2: Witnessed laser phase due to unperturbed energy}
By substituting Eq.~(\ref{Gamma0}) into Eq.~(\ref{odes2}), the witnessed laser phase, $\phi^0$, due to the unperturbed betatron motion reads
\begin{equation}
	\label{laserphase}
	\begin{split}
		&\int_{\phi_0^0}^{\phi^0}d\phi=\int_{\theta_0}^\theta\frac{C_0(1-h^2\cos^2\theta')}{\sqrt{k_e}\sqrt{(\Gamma_{0}^0)^2-C_0h^2+C_0h^2\sin^2\theta'}}d\theta'\\
		&=\frac{1}{\sqrt{k_eg}}\{g[E(q)+E(\theta|q)]+C^\prime[F(q)+F(\theta|q)]\},
	\end{split}
\end{equation}
where $\phi_0^0=\phi^0(\theta=\theta_0)$ and $C^\prime=C_0-(\Gamma_{0}^0)^2$. $F(\theta|q)$ and $E(\theta|q)$ are the first and second kind incomplete elliptic integrals, respectively. $F(q)$ and $E(q)$ are corresponding complete elliptic integrals at $\theta=\pi/2$. $g=(\Gamma_0^0)^2- C_0 h^2, q = - \frac{C_0 h^2}{g}$. Thus, $\phi^0$ takes the form of 
\begin{equation}
	\label{laserphase2}
	\phi^0=\frac{1}{\sqrt{k_eg}}\{g[E(q)+E(\theta|q)]+C^\prime[F(q)+F(\theta|q)]\}+\phi_0^0,
\end{equation}
and it can be further arranged as
\begin{equation}
	\label{laserphase3}
	\begin{split}
		\phi^0&=\phi'+\phi'',\\
		\phi'&=aE(\theta|q)+bF(\theta|q),\\
		\phi''&=aE(q)+bF(q)+\phi_0^0,
	\end{split}
\end{equation}
where $a=\sqrt{g/k_e}, b=C'/\sqrt{k_eg}$.
Making use of a Fourier series expansion of the incomplete elliptic integrals~\cite{cvijovic2010fourier}, $Z(\theta|q)=\frac{2}{\pi}\theta Z(q)+\frac{2}{\pi}\sum_{n\geq 1}I_n^\pm (q)\sin(2n\theta)$ where $Z=E, F$, one has
\begin{equation}
	\label{laserphase4}
		\phi'=l\theta+\sigma\\
\end{equation}
where
\begin{equation}
	\begin{split}
		l&=\frac{2}{\pi}[aE(q)+bF(q)],\\
		\sigma&=\frac{2}{\pi}\sum_{n\geq 1}[aI_n^+(q)+bI_n^-(q)]\sin(2n\theta)=\sum_{n\geq 1}d_n\sin(2n\theta), \\
		d_n&=\frac{2}{\pi}[aI_n^+(q)+bI_n^-(q)].
	\end{split}
\end{equation}

\subsection{Step 3: Work done by laser due to the phase variation}

\begin{figure*}
	\centering
	\includegraphics[width=.7\linewidth]{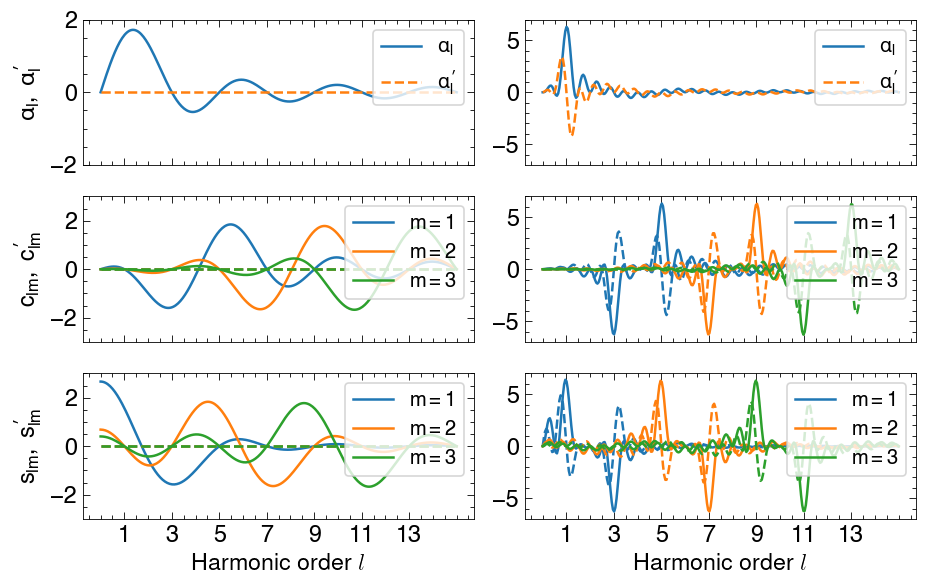}
	\caption{The coefficients of Eq.~(\ref{coefficients}), $\alpha_l'$, $c_{lm}'$, $s_{lm}'$, $\alpha_l$, $c_{lm}$, and $s_{lm}$, obtained by integrating up to (left panel) $\theta'=\pi/2$ and (right panel) $\theta'=7\pi/2$. The dashed lines refer to the $\alpha_l'$, $c_{lm}'$, and the solid lines refer to $\alpha_l$, $c_{lm}, s_{lm}$.}
	\label{coeffs}
\end{figure*}

Upon evaluating $d_n$, the first term ($d_1$) dominates. Therefore we take $\sigma=d_1\sin2\theta$. Now let us plug the laser phase [Eq.~(\ref{laserphase3})] back to $W_L=a_0 h y_b \int_{\theta_0}^{\theta'}  \sin \theta \cos\phi^{0}  d\theta$ to calculate the laser work, which can be arranged as
\begin{equation}
	\begin{split}
		&W_L \ 
		=\chi \frac{\cos\phi''}{2} \int_{\pi/2}^{\theta'}  [S_{l\theta}\cos (d_1\sin 2\theta)+C_{l\theta}\sin (d_1\sin2\theta)] d\theta\\
		&+\chi \frac{\sin\phi''}{2} \int_{\theta_0}^{\theta'}  [C_{l\theta}\cos (d_1\sin2\theta)-S_{l\theta}\sin (d_1\sin2\theta)] d\theta
	\end{split}
\end{equation}
where $\chi=a_0 h y_b, S_{l\theta}=\sin(l+1)\theta-\sin(l-1)\theta, C_{l\theta}=\cos(l+1)\theta-\cos(l-1)\theta$. Further making use of the Jacobi expansion $\cos(z\sin\theta)=J_0(z)+2\sum_{m\geq 1}J_{2m}(z)\cos2m\theta, \sin(z\sin\theta)=2\sum_{m\geq 1}J_{2m-1}(z)\sin(2m-1)\theta$, the integral can be expanded as 
\begin{equation}
	\label{final_WL}
	\begin{split}
		&W_L= \chi\cos \phi'' \left( J_0(d_1) \alpha_l'+ \sum_{m\ge1} \left[ J_{2m}(d_1) c_{lm}' + J_{2m-1}(d_1) s_{lm}' \right] \right)\\
		&- \chi\sin \phi'' \left( J_0(d_1) \alpha_l+ \sum_{m\ge1} \left[ J_{2m}(d_1) c_{lm} + J_{2m-1}(d_1) s_{lm} \right] \right)
	\end{split}
\end{equation}
where the coefficients are 

\begin{equation}
	\label{coefficients}
	\begin{split}
		\alpha_l' &= \frac{1}{2} \int_{\pi/2}^{\theta'} S_{l\theta} d\theta\\
		c_{lm}' &= \int_{\pi/2}^{\theta'} \cos 4m\theta S_{l\theta} d\theta\\
		s_{lm}' &= \int_{\pi/2}^{\theta'} \sin (4m-2)\theta C_{l\theta} d\theta\\
		\alpha_l &= -\frac{1}{2} \int_{-\pi/2}^{\theta'} C_{l\theta} d\theta\\
		c_{lm} &= -\int_{-\pi/2}^{\theta'} \cos 4m\theta C_{l\theta} d\theta\\
		s_{lm} &= \int_{-\pi/2}^{\theta'} \sin (4m-2)\theta S_{l\theta}  d\theta
	\end{split}
\end{equation}

As shown in Fig.~\ref{coeffs}, the variation of these coefficients against $l$ is largely smooth when $\theta'$ is small (i.e., the laser-electron interaction persists over a fraction of the betatron cycle), but quickly narrowed down around odd integers when $\theta'$ increases (i.e., the interaction persists over several betatron cycles). This proves that the odd harmonic resonances ($l=1,3,5,...$) are required for DLA to be effective over several betatron cycles. 

\bibliography{Ref_dla}

\begin{thebibliography}{32}%
\makeatletter
\providecommand \@ifxundefined [1]{%
 \@ifx{#1\undefined}
}%
\providecommand \@ifnum [1]{%
 \ifnum #1\expandafter \@firstoftwo
 \else \expandafter \@secondoftwo
 \fi
}%
\providecommand \@ifx [1]{%
 \ifx #1\expandafter \@firstoftwo
 \else \expandafter \@secondoftwo
 \fi
}%
\providecommand \natexlab [1]{#1}%
\providecommand \enquote  [1]{``#1''}%
\providecommand \bibnamefont  [1]{#1}%
\providecommand \bibfnamefont [1]{#1}%
\providecommand \citenamefont [1]{#1}%
\providecommand \href@noop [0]{\@secondoftwo}%
\providecommand \href [0]{\begingroup \@sanitize@url \@href}%
\providecommand \@href[1]{\@@startlink{#1}\@@href}%
\providecommand \@@href[1]{\endgroup#1\@@endlink}%
\providecommand \@sanitize@url [0]{\catcode `\\12\catcode `\$12\catcode
  `\&12\catcode `\#12\catcode `\^12\catcode `\_12\catcode `\%12\relax}%
\providecommand \@@startlink[1]{}%
\providecommand \@@endlink[0]{}%
\providecommand \url  [0]{\begingroup\@sanitize@url \@url }%
\providecommand \@url [1]{\endgroup\@href {#1}{\urlprefix }}%
\providecommand \urlprefix  [0]{URL }%
\providecommand \Eprint [0]{\href }%
\providecommand \doibase [0]{http://dx.doi.org/}%
\providecommand \selectlanguage [0]{\@gobble}%
\providecommand \bibinfo  [0]{\@secondoftwo}%
\providecommand \bibfield  [0]{\@secondoftwo}%
\providecommand \translation [1]{[#1]}%
\providecommand \BibitemOpen [0]{}%
\providecommand \bibitemStop [0]{}%
\providecommand \bibitemNoStop [0]{.\EOS\space}%
\providecommand \EOS [0]{\spacefactor3000\relax}%
\providecommand \BibitemShut  [1]{\csname bibitem#1\endcsname}%
\let\auto@bib@innerbib\@empty
\bibitem [{\citenamefont {Mourou}\ \emph {et~al.}(2006)\citenamefont {Mourou},
  \citenamefont {Tajima},\ and\ \citenamefont {Bulanov}}]{mourou2006optics}%
  \BibitemOpen
  \bibfield  {author} {\bibinfo {author} {\bibfnamefont {G.~A.}\ \bibnamefont
  {Mourou}}, \bibinfo {author} {\bibfnamefont {T.}~\bibnamefont {Tajima}}, \
  and\ \bibinfo {author} {\bibfnamefont {S.~V.}\ \bibnamefont {Bulanov}},\
  }\href@noop {} {\bibfield  {journal} {\bibinfo  {journal} {Reviews of modern
  physics}\ }\textbf {\bibinfo {volume} {78}},\ \bibinfo {pages} {309}
  (\bibinfo {year} {2006})}\BibitemShut {NoStop}%
\bibitem [{\citenamefont {Esarey}\ \emph {et~al.}(2009)\citenamefont {Esarey},
  \citenamefont {Schroeder},\ and\ \citenamefont
  {Leemans}}]{esarey2009physics}%
  \BibitemOpen
  \bibfield  {author} {\bibinfo {author} {\bibfnamefont {E.}~\bibnamefont
  {Esarey}}, \bibinfo {author} {\bibfnamefont {C.}~\bibnamefont {Schroeder}}, \
  and\ \bibinfo {author} {\bibfnamefont {W.}~\bibnamefont {Leemans}},\
  }\href@noop {} {\bibfield  {journal} {\bibinfo  {journal} {Reviews of modern
  physics}\ }\textbf {\bibinfo {volume} {81}},\ \bibinfo {pages} {1229}
  (\bibinfo {year} {2009})}\BibitemShut {NoStop}%
\bibitem [{\citenamefont {Pukhov}\ \emph {et~al.}(1999)\citenamefont {Pukhov},
  \citenamefont {Sheng},\ and\ \citenamefont {Meyer-ter
  Vehn}}]{pukhov1999particle}%
  \BibitemOpen
  \bibfield  {author} {\bibinfo {author} {\bibfnamefont {A.}~\bibnamefont
  {Pukhov}}, \bibinfo {author} {\bibfnamefont {Z.-M.}\ \bibnamefont {Sheng}}, \
  and\ \bibinfo {author} {\bibfnamefont {J.}~\bibnamefont {Meyer-ter Vehn}},\
  }\href@noop {} {\bibfield  {journal} {\bibinfo  {journal} {Physics of
  Plasmas}\ }\textbf {\bibinfo {volume} {6}},\ \bibinfo {pages} {2847}
  (\bibinfo {year} {1999})}\BibitemShut {NoStop}%
\bibitem [{\citenamefont {Tajima}\ and\ \citenamefont
  {Dawson}(1979)}]{tajima1979laser}%
  \BibitemOpen
  \bibfield  {author} {\bibinfo {author} {\bibfnamefont {T.}~\bibnamefont
  {Tajima}}\ and\ \bibinfo {author} {\bibfnamefont {J.~M.}\ \bibnamefont
  {Dawson}},\ }\href@noop {} {\bibfield  {journal} {\bibinfo  {journal}
  {Physical Review Letters}\ }\textbf {\bibinfo {volume} {43}},\ \bibinfo
  {pages} {267} (\bibinfo {year} {1979})}\BibitemShut {NoStop}%
\bibitem [{\citenamefont {Esarey}\ \emph {et~al.}(1997)\citenamefont {Esarey},
  \citenamefont {Hubbard}, \citenamefont {Leemans}, \citenamefont {Ting},\ and\
  \citenamefont {Sprangle}}]{esarey1997electron}%
  \BibitemOpen
  \bibfield  {author} {\bibinfo {author} {\bibfnamefont {E.}~\bibnamefont
  {Esarey}}, \bibinfo {author} {\bibfnamefont {R.}~\bibnamefont {Hubbard}},
  \bibinfo {author} {\bibfnamefont {W.}~\bibnamefont {Leemans}}, \bibinfo
  {author} {\bibfnamefont {A.}~\bibnamefont {Ting}}, \ and\ \bibinfo {author}
  {\bibfnamefont {P.}~\bibnamefont {Sprangle}},\ }\href@noop {} {\bibfield
  {journal} {\bibinfo  {journal} {Physical Review Letters}\ }\textbf {\bibinfo
  {volume} {79}},\ \bibinfo {pages} {2682} (\bibinfo {year}
  {1997})}\BibitemShut {NoStop}%
\bibitem [{\citenamefont {Pak}\ \emph {et~al.}(2010)\citenamefont {Pak},
  \citenamefont {Marsh}, \citenamefont {Martins}, \citenamefont {Lu},
  \citenamefont {Mori},\ and\ \citenamefont {Joshi}}]{pak2010injection}%
  \BibitemOpen
  \bibfield  {author} {\bibinfo {author} {\bibfnamefont {A.}~\bibnamefont
  {Pak}}, \bibinfo {author} {\bibfnamefont {K.}~\bibnamefont {Marsh}}, \bibinfo
  {author} {\bibfnamefont {S.}~\bibnamefont {Martins}}, \bibinfo {author}
  {\bibfnamefont {W.}~\bibnamefont {Lu}}, \bibinfo {author} {\bibfnamefont
  {W.}~\bibnamefont {Mori}}, \ and\ \bibinfo {author} {\bibfnamefont
  {C.}~\bibnamefont {Joshi}},\ }\href@noop {} {\bibfield  {journal} {\bibinfo
  {journal} {Physical Review Letters}\ }\textbf {\bibinfo {volume} {104}},\
  \bibinfo {pages} {025003} (\bibinfo {year} {2010})}\BibitemShut {NoStop}%
\bibitem [{\citenamefont {Gonsalves}\ \emph {et~al.}(2011)\citenamefont
  {Gonsalves}, \citenamefont {Nakamura}, \citenamefont {Lin}, \citenamefont
  {Panasenko}, \citenamefont {Shiraishi}, \citenamefont {Sokollik},
  \citenamefont {Benedetti}, \citenamefont {Schroeder}, \citenamefont {Geddes},
  \citenamefont {Van~Tilborg} \emph {et~al.}}]{gonsalves2011tunable}%
  \BibitemOpen
  \bibfield  {author} {\bibinfo {author} {\bibfnamefont {A.}~\bibnamefont
  {Gonsalves}}, \bibinfo {author} {\bibfnamefont {K.}~\bibnamefont {Nakamura}},
  \bibinfo {author} {\bibfnamefont {C.}~\bibnamefont {Lin}}, \bibinfo {author}
  {\bibfnamefont {D.}~\bibnamefont {Panasenko}}, \bibinfo {author}
  {\bibfnamefont {S.}~\bibnamefont {Shiraishi}}, \bibinfo {author}
  {\bibfnamefont {T.}~\bibnamefont {Sokollik}}, \bibinfo {author}
  {\bibfnamefont {C.}~\bibnamefont {Benedetti}}, \bibinfo {author}
  {\bibfnamefont {C.}~\bibnamefont {Schroeder}}, \bibinfo {author}
  {\bibfnamefont {C.}~\bibnamefont {Geddes}}, \bibinfo {author} {\bibfnamefont
  {J.}~\bibnamefont {Van~Tilborg}},  \emph {et~al.},\ }\href@noop {} {\bibfield
   {journal} {\bibinfo  {journal} {Nature Physics}\ }\textbf {\bibinfo {volume}
  {7}},\ \bibinfo {pages} {862} (\bibinfo {year} {2011})}\BibitemShut {NoStop}%
\bibitem [{\citenamefont {Gonsalves}\ \emph {et~al.}(2019)\citenamefont
  {Gonsalves}, \citenamefont {Nakamura}, \citenamefont {Daniels}, \citenamefont
  {Benedetti}, \citenamefont {Pieronek}, \citenamefont {De~Raadt},
  \citenamefont {Steinke}, \citenamefont {Bin}, \citenamefont {Bulanov},
  \citenamefont {Van~Tilborg} \emph {et~al.}}]{gonsalves2019petawatt}%
  \BibitemOpen
  \bibfield  {author} {\bibinfo {author} {\bibfnamefont {A.}~\bibnamefont
  {Gonsalves}}, \bibinfo {author} {\bibfnamefont {K.}~\bibnamefont {Nakamura}},
  \bibinfo {author} {\bibfnamefont {J.}~\bibnamefont {Daniels}}, \bibinfo
  {author} {\bibfnamefont {C.}~\bibnamefont {Benedetti}}, \bibinfo {author}
  {\bibfnamefont {C.}~\bibnamefont {Pieronek}}, \bibinfo {author}
  {\bibfnamefont {T.}~\bibnamefont {De~Raadt}}, \bibinfo {author}
  {\bibfnamefont {S.}~\bibnamefont {Steinke}}, \bibinfo {author} {\bibfnamefont
  {J.}~\bibnamefont {Bin}}, \bibinfo {author} {\bibfnamefont {S.}~\bibnamefont
  {Bulanov}}, \bibinfo {author} {\bibfnamefont {J.}~\bibnamefont
  {Van~Tilborg}},  \emph {et~al.},\ }\href@noop {} {\bibfield  {journal}
  {\bibinfo  {journal} {Physical review letters}\ }\textbf {\bibinfo {volume}
  {122}},\ \bibinfo {pages} {084801} (\bibinfo {year} {2019})}\BibitemShut
  {NoStop}%
\bibitem [{\citenamefont {Rosmej}\ \emph {et~al.}(2020)\citenamefont {Rosmej},
  \citenamefont {Gyrdymov}, \citenamefont {G{\"u}nther}, \citenamefont
  {Andreev}, \citenamefont {Tavana}, \citenamefont {Neumayer}, \citenamefont
  {Z{\"a}hter}, \citenamefont {Zahn}, \citenamefont {Popov}, \citenamefont
  {Borisenko} \emph {et~al.}}]{rosmej2020high}%
  \BibitemOpen
  \bibfield  {author} {\bibinfo {author} {\bibfnamefont {O.}~\bibnamefont
  {Rosmej}}, \bibinfo {author} {\bibfnamefont {M.}~\bibnamefont {Gyrdymov}},
  \bibinfo {author} {\bibfnamefont {M.}~\bibnamefont {G{\"u}nther}}, \bibinfo
  {author} {\bibfnamefont {N.}~\bibnamefont {Andreev}}, \bibinfo {author}
  {\bibfnamefont {P.}~\bibnamefont {Tavana}}, \bibinfo {author} {\bibfnamefont
  {P.}~\bibnamefont {Neumayer}}, \bibinfo {author} {\bibfnamefont
  {S.}~\bibnamefont {Z{\"a}hter}}, \bibinfo {author} {\bibfnamefont
  {N.}~\bibnamefont {Zahn}}, \bibinfo {author} {\bibfnamefont {V.}~\bibnamefont
  {Popov}}, \bibinfo {author} {\bibfnamefont {N.}~\bibnamefont {Borisenko}},
  \emph {et~al.},\ }\href@noop {} {\bibfield  {journal} {\bibinfo  {journal}
  {Plasma Physics and Controlled Fusion}\ }\textbf {\bibinfo {volume} {62}},\
  \bibinfo {pages} {115024} (\bibinfo {year} {2020})}\BibitemShut {NoStop}%
\bibitem [{\citenamefont {Kneip}\ \emph {et~al.}(2008)\citenamefont {Kneip},
  \citenamefont {Nagel}, \citenamefont {Bellei}, \citenamefont {Bourgeois},
  \citenamefont {Dangor}, \citenamefont {Gopal}, \citenamefont {Heathcote},
  \citenamefont {Mangles}, \citenamefont {Marques}, \citenamefont {Maksimchuk}
  \emph {et~al.}}]{kneip2008observation}%
  \BibitemOpen
  \bibfield  {author} {\bibinfo {author} {\bibfnamefont {S.}~\bibnamefont
  {Kneip}}, \bibinfo {author} {\bibfnamefont {S.}~\bibnamefont {Nagel}},
  \bibinfo {author} {\bibfnamefont {C.}~\bibnamefont {Bellei}}, \bibinfo
  {author} {\bibfnamefont {N.}~\bibnamefont {Bourgeois}}, \bibinfo {author}
  {\bibfnamefont {A.}~\bibnamefont {Dangor}}, \bibinfo {author} {\bibfnamefont
  {A.}~\bibnamefont {Gopal}}, \bibinfo {author} {\bibfnamefont
  {R.}~\bibnamefont {Heathcote}}, \bibinfo {author} {\bibfnamefont
  {S.}~\bibnamefont {Mangles}}, \bibinfo {author} {\bibfnamefont
  {J.}~\bibnamefont {Marques}}, \bibinfo {author} {\bibfnamefont
  {A.}~\bibnamefont {Maksimchuk}},  \emph {et~al.},\ }\href@noop {} {\bibfield
  {journal} {\bibinfo  {journal} {Physical review letters}\ }\textbf {\bibinfo
  {volume} {100}},\ \bibinfo {pages} {105006} (\bibinfo {year}
  {2008})}\BibitemShut {NoStop}%
\bibitem [{\citenamefont {Liu}\ \emph {et~al.}(2015)\citenamefont {Liu},
  \citenamefont {Hu}, \citenamefont {Wang}, \citenamefont {Wu}, \citenamefont
  {Liu}, \citenamefont {Chen}, \citenamefont {Meyer-ter Vehn}, \citenamefont
  {Yan},\ and\ \citenamefont {He}}]{liu2015quasimonoenergetic}%
  \BibitemOpen
  \bibfield  {author} {\bibinfo {author} {\bibfnamefont {B.}~\bibnamefont
  {Liu}}, \bibinfo {author} {\bibfnamefont {R.}~\bibnamefont {Hu}}, \bibinfo
  {author} {\bibfnamefont {H.}~\bibnamefont {Wang}}, \bibinfo {author}
  {\bibfnamefont {D.}~\bibnamefont {Wu}}, \bibinfo {author} {\bibfnamefont
  {J.}~\bibnamefont {Liu}}, \bibinfo {author} {\bibfnamefont {C.}~\bibnamefont
  {Chen}}, \bibinfo {author} {\bibfnamefont {J.}~\bibnamefont {Meyer-ter
  Vehn}}, \bibinfo {author} {\bibfnamefont {X.}~\bibnamefont {Yan}}, \ and\
  \bibinfo {author} {\bibfnamefont {X.}~\bibnamefont {He}},\ }\href@noop {}
  {\bibfield  {journal} {\bibinfo  {journal} {Physics of Plasmas}\ }\textbf
  {\bibinfo {volume} {22}},\ \bibinfo {pages} {080704} (\bibinfo {year}
  {2015})}\BibitemShut {NoStop}%
\bibitem [{\citenamefont {Chen}\ \emph {et~al.}(2015)\citenamefont {Chen},
  \citenamefont {Fiuza}, \citenamefont {Link}, \citenamefont {Hazi},
  \citenamefont {Hill}, \citenamefont {Hoarty}, \citenamefont {James},
  \citenamefont {Kerr}, \citenamefont {Meyerhofer}, \citenamefont {Myatt} \emph
  {et~al.}}]{chen2015scaling}%
  \BibitemOpen
  \bibfield  {author} {\bibinfo {author} {\bibfnamefont {H.}~\bibnamefont
  {Chen}}, \bibinfo {author} {\bibfnamefont {F.}~\bibnamefont {Fiuza}},
  \bibinfo {author} {\bibfnamefont {A.}~\bibnamefont {Link}}, \bibinfo {author}
  {\bibfnamefont {A.}~\bibnamefont {Hazi}}, \bibinfo {author} {\bibfnamefont
  {M.}~\bibnamefont {Hill}}, \bibinfo {author} {\bibfnamefont {D.}~\bibnamefont
  {Hoarty}}, \bibinfo {author} {\bibfnamefont {S.}~\bibnamefont {James}},
  \bibinfo {author} {\bibfnamefont {S.}~\bibnamefont {Kerr}}, \bibinfo {author}
  {\bibfnamefont {D.}~\bibnamefont {Meyerhofer}}, \bibinfo {author}
  {\bibfnamefont {J.}~\bibnamefont {Myatt}},  \emph {et~al.},\ }\href@noop {}
  {\bibfield  {journal} {\bibinfo  {journal} {Physical review letters}\
  }\textbf {\bibinfo {volume} {114}},\ \bibinfo {pages} {215001} (\bibinfo
  {year} {2015})}\BibitemShut {NoStop}%
\bibitem [{\citenamefont {Fern{\'a}ndez}\ \emph {et~al.}(2017)\citenamefont
  {Fern{\'a}ndez}, \citenamefont {Cort~Gautier}, \citenamefont {Huang},
  \citenamefont {Palaniyappan}, \citenamefont {Albright}, \citenamefont {Bang},
  \citenamefont {Dyer}, \citenamefont {Favalli}, \citenamefont {Hunter},
  \citenamefont {Mendez} \emph {et~al.}}]{fernandez2017laser}%
  \BibitemOpen
  \bibfield  {author} {\bibinfo {author} {\bibfnamefont {J.~C.}\ \bibnamefont
  {Fern{\'a}ndez}}, \bibinfo {author} {\bibfnamefont {D.}~\bibnamefont
  {Cort~Gautier}}, \bibinfo {author} {\bibfnamefont {C.}~\bibnamefont {Huang}},
  \bibinfo {author} {\bibfnamefont {S.}~\bibnamefont {Palaniyappan}}, \bibinfo
  {author} {\bibfnamefont {B.~J.}\ \bibnamefont {Albright}}, \bibinfo {author}
  {\bibfnamefont {W.}~\bibnamefont {Bang}}, \bibinfo {author} {\bibfnamefont
  {G.}~\bibnamefont {Dyer}}, \bibinfo {author} {\bibfnamefont {A.}~\bibnamefont
  {Favalli}}, \bibinfo {author} {\bibfnamefont {J.~F.}\ \bibnamefont {Hunter}},
  \bibinfo {author} {\bibfnamefont {J.}~\bibnamefont {Mendez}},  \emph
  {et~al.},\ }\href@noop {} {\bibfield  {journal} {\bibinfo  {journal} {Physics
  of plasmas}\ }\textbf {\bibinfo {volume} {24}},\ \bibinfo {pages} {056702}
  (\bibinfo {year} {2017})}\BibitemShut {NoStop}%
\bibitem [{\citenamefont {Bin}\ \emph {et~al.}(2018)\citenamefont {Bin},
  \citenamefont {Yeung}, \citenamefont {Gong}, \citenamefont {Wang},
  \citenamefont {Kreuzer}, \citenamefont {Zhou}, \citenamefont {Streeter},
  \citenamefont {Foster}, \citenamefont {Cousens}, \citenamefont {Dromey} \emph
  {et~al.}}]{bin2018enhanced}%
  \BibitemOpen
  \bibfield  {author} {\bibinfo {author} {\bibfnamefont {J.}~\bibnamefont
  {Bin}}, \bibinfo {author} {\bibfnamefont {M.}~\bibnamefont {Yeung}}, \bibinfo
  {author} {\bibfnamefont {Z.}~\bibnamefont {Gong}}, \bibinfo {author}
  {\bibfnamefont {H.}~\bibnamefont {Wang}}, \bibinfo {author} {\bibfnamefont
  {C.}~\bibnamefont {Kreuzer}}, \bibinfo {author} {\bibfnamefont
  {M.}~\bibnamefont {Zhou}}, \bibinfo {author} {\bibfnamefont {M.}~\bibnamefont
  {Streeter}}, \bibinfo {author} {\bibfnamefont {P.}~\bibnamefont {Foster}},
  \bibinfo {author} {\bibfnamefont {S.}~\bibnamefont {Cousens}}, \bibinfo
  {author} {\bibfnamefont {B.}~\bibnamefont {Dromey}},  \emph {et~al.},\
  }\href@noop {} {\bibfield  {journal} {\bibinfo  {journal} {Physical review
  letters}\ }\textbf {\bibinfo {volume} {120}},\ \bibinfo {pages} {074801}
  (\bibinfo {year} {2018})}\BibitemShut {NoStop}%
\bibitem [{\citenamefont {Gahn}\ \emph {et~al.}(1999)\citenamefont {Gahn},
  \citenamefont {Tsakiris}, \citenamefont {Pukhov}, \citenamefont {Meyer-ter
  Vehn}, \citenamefont {Pretzler}, \citenamefont {Thirolf}, \citenamefont
  {Habs},\ and\ \citenamefont {Witte}}]{gahn1999multi}%
  \BibitemOpen
  \bibfield  {author} {\bibinfo {author} {\bibfnamefont {C.}~\bibnamefont
  {Gahn}}, \bibinfo {author} {\bibfnamefont {G.}~\bibnamefont {Tsakiris}},
  \bibinfo {author} {\bibfnamefont {A.}~\bibnamefont {Pukhov}}, \bibinfo
  {author} {\bibfnamefont {J.}~\bibnamefont {Meyer-ter Vehn}}, \bibinfo
  {author} {\bibfnamefont {G.}~\bibnamefont {Pretzler}}, \bibinfo {author}
  {\bibfnamefont {P.}~\bibnamefont {Thirolf}}, \bibinfo {author} {\bibfnamefont
  {D.}~\bibnamefont {Habs}}, \ and\ \bibinfo {author} {\bibfnamefont
  {K.}~\bibnamefont {Witte}},\ }\href@noop {} {\bibfield  {journal} {\bibinfo
  {journal} {Physical Review Letters}\ }\textbf {\bibinfo {volume} {83}},\
  \bibinfo {pages} {4772} (\bibinfo {year} {1999})}\BibitemShut {NoStop}%
\bibitem [{\citenamefont {Liu}\ \emph {et~al.}(2013)\citenamefont {Liu},
  \citenamefont {Wang}, \citenamefont {Liu}, \citenamefont {Fu}, \citenamefont
  {Xu}, \citenamefont {Yan},\ and\ \citenamefont {He}}]{liu2013generating}%
  \BibitemOpen
  \bibfield  {author} {\bibinfo {author} {\bibfnamefont {B.}~\bibnamefont
  {Liu}}, \bibinfo {author} {\bibfnamefont {H.}~\bibnamefont {Wang}}, \bibinfo
  {author} {\bibfnamefont {J.}~\bibnamefont {Liu}}, \bibinfo {author}
  {\bibfnamefont {L.}~\bibnamefont {Fu}}, \bibinfo {author} {\bibfnamefont
  {Y.}~\bibnamefont {Xu}}, \bibinfo {author} {\bibfnamefont {X.}~\bibnamefont
  {Yan}}, \ and\ \bibinfo {author} {\bibfnamefont {X.}~\bibnamefont {He}},\
  }\href@noop {} {\bibfield  {journal} {\bibinfo  {journal} {Physical review
  letters}\ }\textbf {\bibinfo {volume} {110}},\ \bibinfo {pages} {045002}
  (\bibinfo {year} {2013})}\BibitemShut {NoStop}%
\bibitem [{\citenamefont {Palaniyappan}\ \emph {et~al.}(2012)\citenamefont
  {Palaniyappan}, \citenamefont {Hegelich}, \citenamefont {Wu}, \citenamefont
  {Jung}, \citenamefont {Gautier}, \citenamefont {Yin}, \citenamefont
  {Albright}, \citenamefont {Johnson}, \citenamefont {Shimada}, \citenamefont
  {Letzring} \emph {et~al.}}]{palaniyappan2012dynamics}%
  \BibitemOpen
  \bibfield  {author} {\bibinfo {author} {\bibfnamefont {S.}~\bibnamefont
  {Palaniyappan}}, \bibinfo {author} {\bibfnamefont {B.~M.}\ \bibnamefont
  {Hegelich}}, \bibinfo {author} {\bibfnamefont {H.-C.}\ \bibnamefont {Wu}},
  \bibinfo {author} {\bibfnamefont {D.}~\bibnamefont {Jung}}, \bibinfo {author}
  {\bibfnamefont {D.~C.}\ \bibnamefont {Gautier}}, \bibinfo {author}
  {\bibfnamefont {L.}~\bibnamefont {Yin}}, \bibinfo {author} {\bibfnamefont
  {B.~J.}\ \bibnamefont {Albright}}, \bibinfo {author} {\bibfnamefont {R.~P.}\
  \bibnamefont {Johnson}}, \bibinfo {author} {\bibfnamefont {T.}~\bibnamefont
  {Shimada}}, \bibinfo {author} {\bibfnamefont {S.}~\bibnamefont {Letzring}},
  \emph {et~al.},\ }\href@noop {} {\bibfield  {journal} {\bibinfo  {journal}
  {Nature Physics}\ }\textbf {\bibinfo {volume} {8}},\ \bibinfo {pages} {763}
  (\bibinfo {year} {2012})}\BibitemShut {NoStop}%
\bibitem [{\citenamefont {Tsymbalov}\ \emph {et~al.}(2019)\citenamefont
  {Tsymbalov}, \citenamefont {Gorlova}, \citenamefont {Shulyapov},
  \citenamefont {Prokudin}, \citenamefont {Zavorotny}, \citenamefont {Ivanov},
  \citenamefont {Volkov}, \citenamefont {Bychenkov}, \citenamefont {Nedorezov},
  \citenamefont {Paskhalov} \emph {et~al.}}]{tsymbalov2019well}%
  \BibitemOpen
  \bibfield  {author} {\bibinfo {author} {\bibfnamefont {I.}~\bibnamefont
  {Tsymbalov}}, \bibinfo {author} {\bibfnamefont {D.}~\bibnamefont {Gorlova}},
  \bibinfo {author} {\bibfnamefont {S.}~\bibnamefont {Shulyapov}}, \bibinfo
  {author} {\bibfnamefont {V.}~\bibnamefont {Prokudin}}, \bibinfo {author}
  {\bibfnamefont {A.}~\bibnamefont {Zavorotny}}, \bibinfo {author}
  {\bibfnamefont {K.}~\bibnamefont {Ivanov}}, \bibinfo {author} {\bibfnamefont
  {R.}~\bibnamefont {Volkov}}, \bibinfo {author} {\bibfnamefont
  {V.}~\bibnamefont {Bychenkov}}, \bibinfo {author} {\bibfnamefont
  {V.}~\bibnamefont {Nedorezov}}, \bibinfo {author} {\bibfnamefont
  {A.}~\bibnamefont {Paskhalov}},  \emph {et~al.},\ }\href@noop {} {\bibfield
  {journal} {\bibinfo  {journal} {Plasma Physics and Controlled Fusion}\
  }\textbf {\bibinfo {volume} {61}},\ \bibinfo {pages} {075016} (\bibinfo
  {year} {2019})}\BibitemShut {NoStop}%
\bibitem [{\citenamefont {Mangles}\ \emph {et~al.}(2005)\citenamefont
  {Mangles}, \citenamefont {Walton}, \citenamefont {Tzoufras}, \citenamefont
  {Najmudin}, \citenamefont {Clarke}, \citenamefont {Dangor}, \citenamefont
  {Evans}, \citenamefont {Fritzler}, \citenamefont {Gopal}, \citenamefont
  {Hernandez-Gomez} \emph {et~al.}}]{mangles2005electron}%
  \BibitemOpen
  \bibfield  {author} {\bibinfo {author} {\bibfnamefont {S.~P.}\ \bibnamefont
  {Mangles}}, \bibinfo {author} {\bibfnamefont {B.}~\bibnamefont {Walton}},
  \bibinfo {author} {\bibfnamefont {M.}~\bibnamefont {Tzoufras}}, \bibinfo
  {author} {\bibfnamefont {Z.}~\bibnamefont {Najmudin}}, \bibinfo {author}
  {\bibfnamefont {R.}~\bibnamefont {Clarke}}, \bibinfo {author} {\bibfnamefont
  {A.~E.}\ \bibnamefont {Dangor}}, \bibinfo {author} {\bibfnamefont
  {R.}~\bibnamefont {Evans}}, \bibinfo {author} {\bibfnamefont
  {S.}~\bibnamefont {Fritzler}}, \bibinfo {author} {\bibfnamefont
  {A.}~\bibnamefont {Gopal}}, \bibinfo {author} {\bibfnamefont
  {C.}~\bibnamefont {Hernandez-Gomez}},  \emph {et~al.},\ }\href@noop {}
  {\bibfield  {journal} {\bibinfo  {journal} {Physical review letters}\
  }\textbf {\bibinfo {volume} {94}},\ \bibinfo {pages} {245001} (\bibinfo
  {year} {2005})}\BibitemShut {NoStop}%
\bibitem [{\citenamefont {Zhang}\ and\ \citenamefont
  {Krasheninnikov}(2018)}]{zhang2018electron}%
  \BibitemOpen
  \bibfield  {author} {\bibinfo {author} {\bibfnamefont {Y.}~\bibnamefont
  {Zhang}}\ and\ \bibinfo {author} {\bibfnamefont {S.}~\bibnamefont
  {Krasheninnikov}},\ }\href@noop {} {\bibfield  {journal} {\bibinfo  {journal}
  {Physics of Plasmas}\ }\textbf {\bibinfo {volume} {25}},\ \bibinfo {pages}
  {013120} (\bibinfo {year} {2018})}\BibitemShut {NoStop}%
\bibitem [{\citenamefont {Tsakiris}\ \emph {et~al.}(2000)\citenamefont
  {Tsakiris}, \citenamefont {Gahn},\ and\ \citenamefont
  {Tripathi}}]{tsakiris2000laser}%
  \BibitemOpen
  \bibfield  {author} {\bibinfo {author} {\bibfnamefont {G.}~\bibnamefont
  {Tsakiris}}, \bibinfo {author} {\bibfnamefont {C.}~\bibnamefont {Gahn}}, \
  and\ \bibinfo {author} {\bibfnamefont {V.}~\bibnamefont {Tripathi}},\
  }\href@noop {} {\bibfield  {journal} {\bibinfo  {journal} {Physics of
  Plasmas}\ }\textbf {\bibinfo {volume} {7}},\ \bibinfo {pages} {3017}
  (\bibinfo {year} {2000})}\BibitemShut {NoStop}%
\bibitem [{\citenamefont {Khudik}\ \emph {et~al.}(2016)\citenamefont {Khudik},
  \citenamefont {Arefiev}, \citenamefont {Zhang},\ and\ \citenamefont
  {Shvets}}]{khudik2016universal}%
  \BibitemOpen
  \bibfield  {author} {\bibinfo {author} {\bibfnamefont {V.}~\bibnamefont
  {Khudik}}, \bibinfo {author} {\bibfnamefont {A.}~\bibnamefont {Arefiev}},
  \bibinfo {author} {\bibfnamefont {X.}~\bibnamefont {Zhang}}, \ and\ \bibinfo
  {author} {\bibfnamefont {G.}~\bibnamefont {Shvets}},\ }\href@noop {}
  {\bibfield  {journal} {\bibinfo  {journal} {Physics of Plasmas}\ }\textbf
  {\bibinfo {volume} {23}},\ \bibinfo {pages} {103108} (\bibinfo {year}
  {2016})}\BibitemShut {NoStop}%
\bibitem [{\citenamefont {Zhang}\ \emph {et~al.}(2018)\citenamefont {Zhang},
  \citenamefont {Krasheninnikov},\ and\ \citenamefont
  {Knyazev}}]{zhang2018stochastic}%
  \BibitemOpen
  \bibfield  {author} {\bibinfo {author} {\bibfnamefont {Y.}~\bibnamefont
  {Zhang}}, \bibinfo {author} {\bibfnamefont {S.}~\bibnamefont
  {Krasheninnikov}}, \ and\ \bibinfo {author} {\bibfnamefont {A.}~\bibnamefont
  {Knyazev}},\ }\href@noop {} {\bibfield  {journal} {\bibinfo  {journal}
  {Physics of Plasmas}\ }\textbf {\bibinfo {volume} {25}},\ \bibinfo {pages}
  {123110} (\bibinfo {year} {2018})}\BibitemShut {NoStop}%
\bibitem [{\citenamefont {Huang}\ \emph {et~al.}(2017)\citenamefont {Huang},
  \citenamefont {Zhou}, \citenamefont {Robinson}, \citenamefont {Qiao},
  \citenamefont {Arefiev}, \citenamefont {Norreys}, \citenamefont {He},\ and\
  \citenamefont {Ruan}}]{huang2017nonlinear}%
  \BibitemOpen
  \bibfield  {author} {\bibinfo {author} {\bibfnamefont {T.~W.}\ \bibnamefont
  {Huang}}, \bibinfo {author} {\bibfnamefont {C.~T.}\ \bibnamefont {Zhou}},
  \bibinfo {author} {\bibfnamefont {A.~P.~L.}\ \bibnamefont {Robinson}},
  \bibinfo {author} {\bibfnamefont {B.}~\bibnamefont {Qiao}}, \bibinfo {author}
  {\bibfnamefont {A.~V.}\ \bibnamefont {Arefiev}}, \bibinfo {author}
  {\bibfnamefont {P.~A.}\ \bibnamefont {Norreys}}, \bibinfo {author}
  {\bibfnamefont {X.~T.}\ \bibnamefont {He}}, \ and\ \bibinfo {author}
  {\bibfnamefont {S.~C.}\ \bibnamefont {Ruan}},\ }\href@noop {} {\bibfield
  {journal} {\bibinfo  {journal} {Physics of Plasmas}\ }\textbf {\bibinfo
  {volume} {24}},\ \bibinfo {pages} {043105} (\bibinfo {year}
  {2017})}\BibitemShut {NoStop}%
\bibitem [{\citenamefont {Arefiev}\ \emph {et~al.}(2012)\citenamefont
  {Arefiev}, \citenamefont {Breizman}, \citenamefont {Schollmeier},\ and\
  \citenamefont {Khudik}}]{arefiev2012parametric}%
  \BibitemOpen
  \bibfield  {author} {\bibinfo {author} {\bibfnamefont {A.~V.}\ \bibnamefont
  {Arefiev}}, \bibinfo {author} {\bibfnamefont {B.~N.}\ \bibnamefont
  {Breizman}}, \bibinfo {author} {\bibfnamefont {M.}~\bibnamefont
  {Schollmeier}}, \ and\ \bibinfo {author} {\bibfnamefont {V.~N.}\ \bibnamefont
  {Khudik}},\ }\href@noop {} {\bibfield  {journal} {\bibinfo  {journal}
  {Physical review letters}\ }\textbf {\bibinfo {volume} {108}},\ \bibinfo
  {pages} {145004} (\bibinfo {year} {2012})}\BibitemShut {NoStop}%
\bibitem [{\citenamefont {Meyer-ter Vehn}\ \emph {et~al.}(2001)\citenamefont
  {Meyer-ter Vehn}, \citenamefont {Pukhov},\ and\ \citenamefont
  {Sheng}}]{meyer2001relativistic}%
  \BibitemOpen
  \bibfield  {author} {\bibinfo {author} {\bibfnamefont {J.}~\bibnamefont
  {Meyer-ter Vehn}}, \bibinfo {author} {\bibfnamefont {A.}~\bibnamefont
  {Pukhov}}, \ and\ \bibinfo {author} {\bibfnamefont {Z.-M.}\ \bibnamefont
  {Sheng}},\ }in\ \href@noop {} {\emph {\bibinfo {booktitle} {Atoms, Solids,
  and Plasmas in Super-Intense Laser Fields}}}\ (\bibinfo  {publisher}
  {Springer},\ \bibinfo {year} {2001})\ pp.\ \bibinfo {pages}
  {167--192}\BibitemShut {NoStop}%
\bibitem [{\citenamefont {Zhang}\ \emph {et~al.}(2015)\citenamefont {Zhang},
  \citenamefont {Khudik},\ and\ \citenamefont {Shvets}}]{zhang2015synergistic}%
  \BibitemOpen
  \bibfield  {author} {\bibinfo {author} {\bibfnamefont {X.}~\bibnamefont
  {Zhang}}, \bibinfo {author} {\bibfnamefont {V.~N.}\ \bibnamefont {Khudik}}, \
  and\ \bibinfo {author} {\bibfnamefont {G.}~\bibnamefont {Shvets}},\
  }\href@noop {} {\bibfield  {journal} {\bibinfo  {journal} {Physical review
  letters}\ }\textbf {\bibinfo {volume} {114}},\ \bibinfo {pages} {184801}
  (\bibinfo {year} {2015})}\BibitemShut {NoStop}%
\bibitem [{\citenamefont {Downer}\ \emph {et~al.}(2018)\citenamefont {Downer},
  \citenamefont {Zgadzaj}, \citenamefont {Debus}, \citenamefont {Schramm},\
  and\ \citenamefont {Kaluza}}]{downer2018diagnostics}%
  \BibitemOpen
  \bibfield  {author} {\bibinfo {author} {\bibfnamefont {M.}~\bibnamefont
  {Downer}}, \bibinfo {author} {\bibfnamefont {R.}~\bibnamefont {Zgadzaj}},
  \bibinfo {author} {\bibfnamefont {A.}~\bibnamefont {Debus}}, \bibinfo
  {author} {\bibfnamefont {U.}~\bibnamefont {Schramm}}, \ and\ \bibinfo
  {author} {\bibfnamefont {M.}~\bibnamefont {Kaluza}},\ }\href@noop {}
  {\bibfield  {journal} {\bibinfo  {journal} {Reviews of Modern Physics}\
  }\textbf {\bibinfo {volume} {90}},\ \bibinfo {pages} {035002} (\bibinfo
  {year} {2018})}\BibitemShut {NoStop}%
\bibitem [{\citenamefont {Huang}\ and\ \citenamefont
  {Kim}(2007)}]{huang2007review}%
  \BibitemOpen
  \bibfield  {author} {\bibinfo {author} {\bibfnamefont {Z.}~\bibnamefont
  {Huang}}\ and\ \bibinfo {author} {\bibfnamefont {K.-J.}\ \bibnamefont
  {Kim}},\ }\href@noop {} {\bibfield  {journal} {\bibinfo  {journal} {Physical
  Review Special Topics-Accelerators and Beams}\ }\textbf {\bibinfo {volume}
  {10}},\ \bibinfo {pages} {034801} (\bibinfo {year} {2007})}\BibitemShut
  {NoStop}%
\bibitem [{\citenamefont {Dunning}\ \emph {et~al.}(2013)\citenamefont
  {Dunning}, \citenamefont {Hemsing}, \citenamefont {Hast}, \citenamefont
  {Raubenheimer}, \citenamefont {Weathersby}, \citenamefont {Xiang},\ and\
  \citenamefont {Fu}}]{dunning2013demonstration}%
  \BibitemOpen
  \bibfield  {author} {\bibinfo {author} {\bibfnamefont {M.}~\bibnamefont
  {Dunning}}, \bibinfo {author} {\bibfnamefont {E.}~\bibnamefont {Hemsing}},
  \bibinfo {author} {\bibfnamefont {C.}~\bibnamefont {Hast}}, \bibinfo {author}
  {\bibfnamefont {T.}~\bibnamefont {Raubenheimer}}, \bibinfo {author}
  {\bibfnamefont {S.}~\bibnamefont {Weathersby}}, \bibinfo {author}
  {\bibfnamefont {D.}~\bibnamefont {Xiang}}, \ and\ \bibinfo {author}
  {\bibfnamefont {F.}~\bibnamefont {Fu}},\ }\href@noop {} {\bibfield  {journal}
  {\bibinfo  {journal} {Physical review letters}\ }\textbf {\bibinfo {volume}
  {110}},\ \bibinfo {pages} {244801} (\bibinfo {year} {2013})}\BibitemShut
  {NoStop}%
\bibitem [{\citenamefont {Wang}\ \emph {et~al.}(2020)\citenamefont {Wang},
  \citenamefont {Ribeyre}, \citenamefont {Gong}, \citenamefont {Jansen},
  \citenamefont {d'Humi{\`e}res}, \citenamefont {Stutman}, \citenamefont
  {Toncian},\ and\ \citenamefont {Arefiev}}]{wang2020power}%
  \BibitemOpen
  \bibfield  {author} {\bibinfo {author} {\bibfnamefont {T.}~\bibnamefont
  {Wang}}, \bibinfo {author} {\bibfnamefont {X.}~\bibnamefont {Ribeyre}},
  \bibinfo {author} {\bibfnamefont {Z.}~\bibnamefont {Gong}}, \bibinfo {author}
  {\bibfnamefont {O.}~\bibnamefont {Jansen}}, \bibinfo {author} {\bibfnamefont
  {E.}~\bibnamefont {d'Humi{\`e}res}}, \bibinfo {author} {\bibfnamefont
  {D.}~\bibnamefont {Stutman}}, \bibinfo {author} {\bibfnamefont
  {T.}~\bibnamefont {Toncian}}, \ and\ \bibinfo {author} {\bibfnamefont
  {A.}~\bibnamefont {Arefiev}},\ }\href@noop {} {\bibfield  {journal} {\bibinfo
   {journal} {Phys. Rev. Applied}\ }\textbf {\bibinfo {volume} {13}},\ \bibinfo
  {pages} {054024} (\bibinfo {year} {2020})}\BibitemShut {NoStop}%
\bibitem [{\citenamefont {Cvijovi{\'c}}(2010)}]{cvijovic2010fourier}%
  \BibitemOpen
  \bibfield  {author} {\bibinfo {author} {\bibfnamefont {D.}~\bibnamefont
  {Cvijovi{\'c}}},\ }\href@noop {} {\bibfield  {journal} {\bibinfo  {journal}
  {Integral Transforms and Special Functions}\ }\textbf {\bibinfo {volume}
  {21}},\ \bibinfo {pages} {235} (\bibinfo {year} {2010})}\BibitemShut
  {NoStop}%
\end{thebibliography}%

\end{document}